\documentclass[11pt,a4paper]{article}
\usepackage[body={16cm,25cm}]{geometry}
\usepackage[dvips]{color}
\usepackage{pgf,pgfarrows}

\usepackage[hang,small,center]{caption}
\usepackage{amsmath,amssymb}
\usepackage{amsfonts}
\usepackage{mathrsfs}
\usepackage{epic}
\usepackage{eepic}
\usepackage{graphicx}
\usepackage{cite}
\newcommand{\ds}{\displaystyle}
\renewcommand{\author}[1]{\large\rm #1\\ \bigskip}
\newcommand{\address}[1]{{\normalsize\it #1\\}\bigskip}
\renewcommand{\title}[1]{\bigskip\bigskip\Large\bf #1\bigskip\bigskip\\}

\newcommand{\Bigpsi}[3]{\phantom{\Psi}_2 \kern -.05em
\Psi_2\left(\genfrac{}{}{0pt}{}{#1}{#2}\biggl|#3\right)}

\def\bea{\begin{eqnarray}}
\def\eea{\end{eqnarray}}
\newcommand{\be}{\begin{equation}}
\newcommand{\ee}{\end{equation}}
\newcommand{\beq}{\begin{equation}}
\newcommand{\eeq}{\end{equation}}
\newcommand{\bedef}{\stackrel{\textrm{def}}{=}}

\renewcommand{\L}{{\mathbf L}}

\newcommand{\Ru}{{\mathbf R}}
\newcommand{\Lu}{{\mathbf L}}
\newcommand{\Su}{{\mathbf S}}
\newcommand{\Sc}{{\mathcal S}}
 
\newcommand{\Gu}{{\mathbf G}} 
\newcommand{\Fu}{{\mathbf F}} 
\newcommand{\ii}{\mathsf{i}}

\newcommand{\C}{{\mathbb C}^2}
\renewcommand{\t}{{\theta}}     
\newcommand{\ts}{{\overline{\theta}}}     

\newcommand{\w}{{\omega}}

\renewcommand{\l}{{\lambda}}

\DeclareMathOperator{\cn}{cn}
\DeclareMathOperator{\sn}{sn}
\DeclareMathOperator{\dn}{dn}
\DeclareMathOperator{\cd}{cd}
\DeclareMathOperator{\sd}{sd}

\newcounter{app}
\newcounter{sapp}[app]

\begin{document}
\vglue 2 cm
\begin{center}
\title{An Elliptic Parameterisation of the Zamolodchikov Model}
\author{Vladimir V.~Bazhanov$^{1,2}$, Vladimir
  V.~Mangazeev$^{1,2}$, Yuichiro Okada$^1$ \\ and \\ Sergey M.~Sergeev$^3$}
\address{$^1$Department of Theoretical Physics,
         Research School of Physics and Engineering,\\
    Australian National University, Canberra, ACT 0200, Australia.\\\ \\
$^2$Mathematical Sciences Institute,\\
      Australian National University, Canberra, ACT 0200,
      Australia.\\\ \\
$^3$Faculty of Information Sciences and Engineering,\\
University of Canberra, Bruce ACT 2601, Australia.}

\vspace{2cm}
{\em Dedicated to Sasha Zamolodchikov on the occasion of his
  sixtieth birthday}
\vspace{1cm}

\begin{abstract} 
The Zamolodchikov model describes an exact relativistic factorized
scattering theory of straight strings in $(2+1)$-dimensional
space-time. It also defines an integrable 3D lattice model of
statistical mechanics and quantum field theory. The three-string $S$-matrix
satisfies the {\em tetrahedron equation} which is a 3D analog of the
Yang-Baxter equation. Each $S$-matrix depends on three dihedral angles
formed by three intersecting planes, whereas the tetrahedron equation
contains five independent spectral parameters, associated with
angles of an Euclidean tetrahedron. The vertex weights are given
by rather complicated expressions involving square roots of 
trigonometric function of the spectral parameters, which is quite
unusual from the point of view of 2D solvable lattice models. In this
paper we consider a particular four-parameter specialization of the
tetrahedron equation when one of its vertices goes to infinity and the 
tetrahedron itself degenerates into an infinite prism. We show that in
this limit all the vertex weights in the tetrahedron equation can be
represented as meromorphic functions on an elliptic curve. Moreover we
show that a special reduction of the tetrahedron equation in this case 
leads precisely to an example
of the tetrahedral Zamolodchikov algebra, previously constructed by
Korepanov. This algebra plays important role for a ``layered'' 
construction of the Shastry's $R$-matrix and the 
2D $S$-matrix appearing in the problem of the
ADS/CFT correspondence for ${\cal N}=4$ SUSY Yang-Mills theory
in four dimensions. 
Possible applications of our results in this field
are briefly discussed.

\end{abstract}

\end{center}

\newpage

\section{Introduction}
The {\em tetrahedron equation} 
\cite{Zamolodchikov:1980rus,Zamolodchikov:1981kf} is a 
three-dimensional analog of the Yang-Baxter equation. It implies
the commutativity of layer-to-layer transfer matrices
\cite{Bazhanov:1981zm} for 
three-dimensional lattice models of statistical mechanics and
field theory and, thus, generalizes the most
fundamental integrability structure of exactly solvable models in two
dimensions \cite{Bax82}.

The first solution of the tetrahedron equation was proposed by Zamolodchikov
\cite{Zamolodchikov:1980rus,Zamolodchikov:1981kf} in the context of an exact 
relativistic factorized scattering theory of ``straight strings''
in 2+1 dimensional space-time. 
Subsequently, Baxter \cite{Baxter:1983qc}
proved that this solution indeed satisfies the tetrahedron
equations and then \cite{Baxter:1986phd}
exactly calculated the free energy of the
corresponding solvable three-dimensional model in the limit of an
infinite lattice. Some further developments in the field of 3D 
integrability in a more general setting could be learned from
\cite{Bazhanov:1992jqa,Bazhanov:1993j,KMS:1993,Maillet:1989gg,Bazhanov:2005as,
  Bazhanov:2008rd}.  
Many special properties of the Zamolodchikov model were studied in
\cite{BaxterForrester:1985,BaxterQuispel:1990,
  BaxterBazhanov:1997,BoosMangazeev:1999}.  
In this paper we will consider some additional analytic and algebraic
problems related to this model. 

The spin variables in the Zamolodchikov model take two values.  In the 
original formulation \cite{Zamolodchikov:1980rus,
  Zamolodchikov:1981kf} the spins are assigned to faces of the
lattice. Baxter \cite{Baxter:1986phd} used a different, but closely
related ``interaction-round-a-cube'' (IRC) approach, 
where the spins are assigned to 3D cells of the
lattice, or, equivalently, vertices of the dual lattice. Here we will
use yet another approach, namely the {\em vertex formulation} with
edge spins, found by Sergeev, Mangazeev and Stroganov
\cite{Sergeev:1995rt}, which is related to the other formulations by 
the ``cube-vertex'' correspondence \cite{Sergeev:1995fb}. The most general
edge-spin solution
of the tetrahedron equation in the
model contains five independent spectral parameters associated with
angles of an Euclidean tetrahedron. Two different reductions of this
solution with a fewer number of parameters, known as the 
``static'' and ``planar'' limits, were previously found by 
Korepanov \cite{Korepanov:1993} and Hietarinta
\cite{Hietarinta:1994}.

In this paper we consider another particular limit when one of the
tetrahedron vertices goes to infinity, 
whereas the tetrahedron turns into an infinite
prism. We show that the all Boltzmann weights in this case can be
parametrised in terms of meromorphic functions on an 
elliptic curve. Moreover we show that a special reduction of the
tetrahedron equation in this case leads precisely to the example of
the Zamolodchikov tetrahedral algebra, constructed by
Korepanov \cite{Korepanov:1993}. This algebra plays an important
role for a ``layered'' construction \cite{ShWad95a,ShWad95b,Hubbook} 
of the Shastry's $R$-matrix \cite{Shastry:1986}
underlying the integrability of the 1D Hubbard model. 
Remarkably, the same $R$-matrix
\cite{Beisert:2006qh,Arutyunov:2006yd, Martins:2007hb,Mitev:2012vt}
appears in the problem of the AdS/CFT correspondence for ${\cal N}=4$
SUSY Yang-Mills theory in four dimensions. In conclusion we briefly
discuss possible applications of our results to this field. 
 
The organization of the paper is as follows. In Sect.2 we briefly
review known results on the tetrahedron equation in the
Zamolodchikov model, used in this paper. In Sect.3 we
derive an elliptic parameterisation of the Boltzmann weights. 
In  Sect.4 we consider special limits of the 
tetrahedron equation and the tetrahedral Zamolodchikov algebra.

\section{The vertex formulation of the Zamolodchikov model} 

\begin{figure}[hbt]
\centering
\includegraphics[height=9cm]{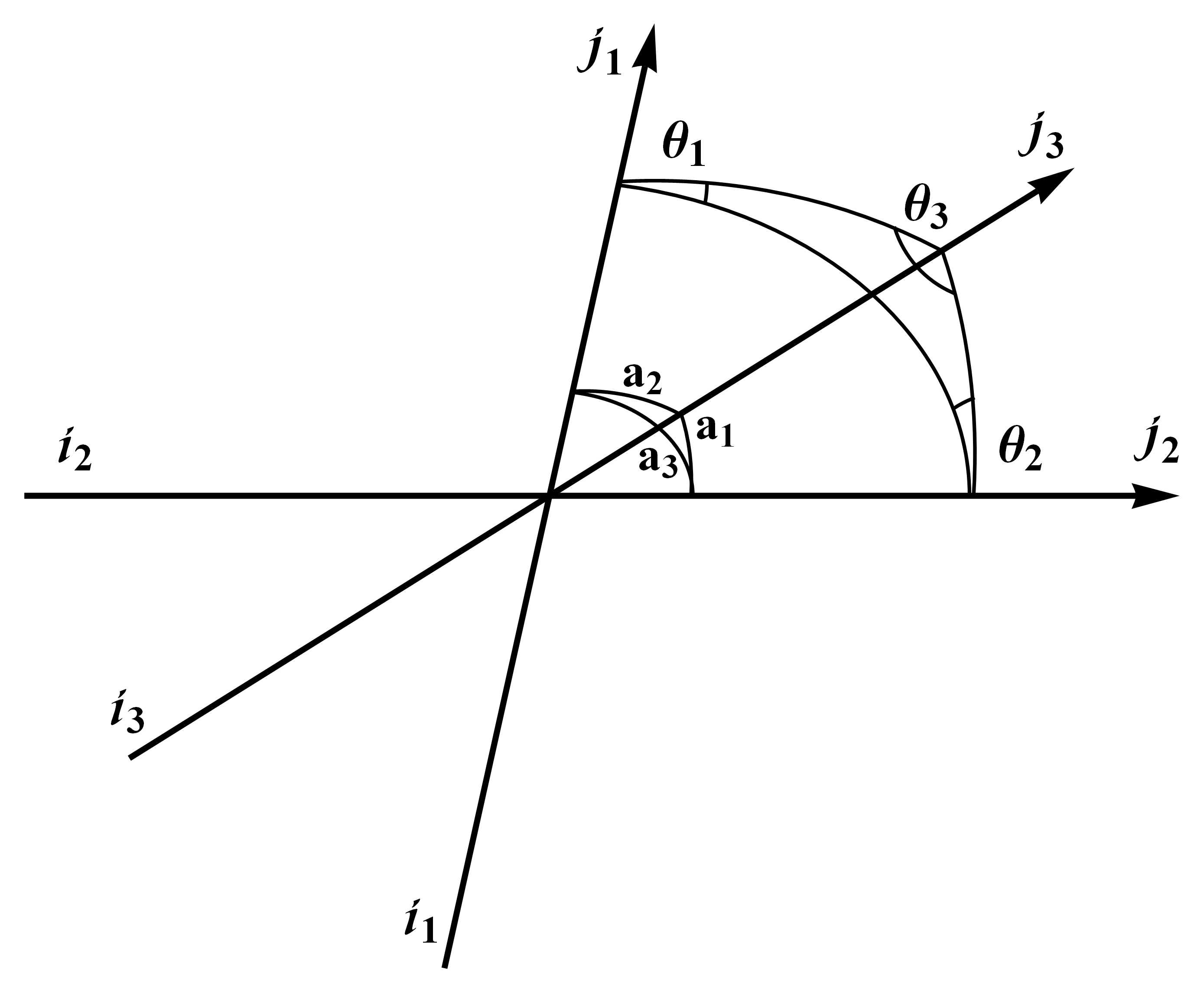}
\caption{
Arrangement of spins around an elementary vertex formed by three
intersecting planes with the dihedral angles $\t_1,\t_2,\t_3$ and the
edge angles $a_1,a_2,a_3$.
}\label{fig1}
\end{figure}

The Zamolodchikov model can be formulated on any 3D lattice formed by
an arbitrary set of intersecting 2D planes in  the Euclidean space 
${\mathbb R}^3$, such that 
no four planes intersect at one point. 
In particular, this definition includes a regular cubic
lattice as the simplest possibility.
The fluctuating spin variables in the model take two
values, denoted below $0$ and $1$. Here we will use the vertex formulation  
of the model found in \cite{Sergeev:1995rt}. 
The spins in this case are assigned to
edges of the lattice while local Boltzmann weights  
are assigned to vertices at three-plane  
intersection points.  
Let $R_{i_1,i_2,i_3}^{j_1,j_2,j_3}$ denote the weight  
corresponding to a configuration of six edge spins
$i_1,i_2,i_3,j_1,j_2,j_3=0,1$, arranged as in Fig.\ref{fig1}.
This quantity could be conveniently
associated with a linear operator 
\beq
\Ru:\qquad   \C\otimes\C\otimes
\C\rightarrow\C\otimes\C\otimes\C\ ,\label{R-def}
\eeq
acting in a direct product of three two-dimensional vector
spaces, where the pairs of indices $(i_1,j_1)$,\  $(i_2,j_2)$ and $(i_3,j_3)$
serve as matrix indices in the first, second and third spaces,
respectively. The edge 
states ``$0$'' and ``$1$'' correspond to the basis vectors $v_0$ and $v_1$,
\beq
v_0=\left( {1 \atop 0} \right),\qquad v_1=\left( {0 \atop 1}
\right),\qquad v_0,v_1\in \C.\label{v01-def}
\eeq 
The vertex weights depends on three spectral parameters, 
$\t_1,\t_2,\t_3$, which are identified with dihedral
angles between the three planes forming the vertex, as shown in
Fig.\ref{fig1}. To indicate this dependence we will write the weights
as 
$\Ru(\t_1,\t_2,\t_3)$. Consider a spherical triangle with the angles 
$\t_1,\t_2,\t_3$ and let $a_1,a_2,a_3$ denote three sides of this
triangle opposite to the angles $\t_1,\t_2,\t_3$, which obey the
spherical sine theorem 
\beq
K=\frac{\sin \t_1}{\sin a_1}=\frac{\sin \t_2}{\sin
  a_2}=\frac{\sin \t_3}{\sin a_3}\,.\label{sine}
\eeq
Define related
variables 
\beq
\begin{array}{rclrcl}
2\,\alpha_0&=&\t_1+\t_2+\t_3-\pi,\qquad &\alpha_i&=&\theta_i-\alpha_0,\\[.3cm]
2\,\beta_0&=&2\pi-a_1-a_2-a_3,\qquad &\beta_i&=&\pi-a_i-\beta_0,
\end{array}\label{al-def}
\eeq
for $i=1,2,3$. They satisfy the relations 
\beq
\alpha_0+\alpha_1+\alpha_2+\alpha_3=\pi,\qquad
\beta_0+\beta_1+\beta_2+\beta_3=\pi\ .\label{alpha-sum}
\eeq
The variables $\alpha_0,\alpha_1,\alpha_2,\alpha_3$ are called {\em excesses} 
of the spherical triangle. 
Choose $\t_1,\t_2,\t_3$ so that $\t_1,\t_2,\t_3$,\ \ \  
$\alpha_0,\ldots,\alpha_3$ and $\beta_0,\ldots,\beta_3$ are all real
and between $0$ and $\pi$. 
Further, define four variables 
(taking positive values of
the square roots)
\beq
t_i=\sqrt{\tan\frac{\alpha_i}{2}},\qquad i=0,1,2,3\ ,\label{t-def}
\eeq
which are constrained by the relation
\begin{equation}\label{const}
1-t_0^2t_1^2-t_0^2t_2^2-t_0^2t_3^2-t_1^2t_2^2-t_1^2t_3^2-t_2^2t_3^2+t_0^2t_1^2t_2^2t_3^2=0,
\end{equation}
following from \eqref{alpha-sum}.
The weights obey a ``parity conservation law'', such that 32 vertex
configurations with an odd number of $0$'s have the identically
vanishing weight   
\beq
R_{i_1i_2i_3}^{j_1j_2j_3}\equiv0, \quad \mbox{if} \quad
i_1+i_2+i_3\not=j_1+j_2+j_3 \pmod 2\ .\label{parity}
\eeq
The remaining 32 non-vanishing 
matrix elements of $\Ru(\t_1,\t_2,\t_3)$ have
the form \cite{Sergeev:1995rt}\footnote{In \cite{Sergeev:1995rt} the weights   
are given in terms of the variables
$\beta_0,\beta_1,\beta_2,\beta_3$. Here they are re-expressed in terms
of $\alpha$'s with the help of the identity
\beq
\tan\frac{\alpha_i}{2}\,\tan\frac{\alpha_j}{2}=
\tan\frac{\beta_k}{2}\, \tan\frac{\beta_\ell}{2} 
\eeq
where $(i,j,k,\ell)$ is any permutation of the
indices $(0,1,2,3)$.
}
\beq
\begin{array}{rcrcrcrcl}
R_{0,0,0}^{0,0,0}&= & R_{0,1,1}^{0,1,1}&= & R_{1,0,1}^{1,0,1}&= & R_{1,1,0}^{1,1,0}&= & 1\\ \\

R_{1,1,1}^{1,1,1}&= & R_{1,0,0}^{1,0,0}&= & R_{0,1,0}^{0,1,0}&= & R_{0,0,1}^{0,0,1}&= & t_0t_1t_2t_3\\ \\
R_{0,0,1}^{0,1,0}&= & R_{0,1,0}^{0,0,1}&= & -R_{1,1,1}^{1,0,0}&= & -R_{1,0,0}^{1,1,1}&= & t_2t_3\\ \\
R_{1,1,0}^{1,0,1}&= & R_{1,0,1}^{1,1,0}&= & -R_{0,0,0}^{0,1,1}&= & -R_{0,1,1}^{0,0,0}&= & t_0t_1\\ \\
R_{0,1,0}^{1,1,1}&= & R_{0,0,1}^{1,0,0}&= & -R_{1,0,0}^{0,0,1}&= & -R_{1,1,1}^{0,1,0}&= & -\ii t_1t_3\\ \\
R_{1,0,1}^{0,0,0}&= & R_{1,1,0}^{0,1,1}&= & -R_{0,1,1}^{1,1,0}&= & -R_{0,0,0}^{1,0,1}&= & \ii t_0t_2\\ \\
R_{1,1,1}^{0,0,1}&= & R_{0,0,1}^{1,1,1}&= & R_{0,1,0}^{1,0,0}&= & R_{1,0,0}^{0,1,0}&= & t_1t_2\\ \\
R_{0,0,0}^{1,1,0}&= & R_{1,1,0}^{0,0,0}&= & R_{1,0,1}^{0,1,1}&= & R_{0,1,1}^{1,0,1}&= & t_0t_3\\ 
\end{array}\label{weights}
\eeq
In the following we will regard the weights 
as functions of the three independent variables
$\t_1,\t_2,\t_3$, 
or, equivalently, of the four dependent variables
$\alpha_0,\alpha_1,\alpha_2,\alpha_3$, 
constrained by the relation \eqref{alpha-sum}.
We would like to stress that the expressions \eqref{weights}
do not contains any
other free parameters. 

In the original face-spin formulation of
ref.\cite{Zamolodchikov:1981kf} the weights are explicitly symmetric   
with respect to all spatial symmetry transformations, generated by
permutations of the lines $1$, $2$, $3$ and inversions of their
directions. Namely, the weights remains unchanged if 
the associated permutations of spins are accompanied by the corresponding
transformations of the dihedral angles.   
Contrary, to this the operator $\Ru(\t_1,\t_2,\t_3)$ is not explicitly
symmetric. Nevertheless it possesses the same spatial symmetry
group of the 3D cube \cite{Sergeev:1995rt,Kashaev:1993ijmp}, but its
realisation now involves linear similarity transformations of the weights.  
This 48-element group is
generated just by two transformations, which we choose as  
\begin{equation}\begin{array}{rccc}
&\Ru(\pi-\t_1,\pi-\t_2,\t_3)&=&
(\sigma_y\otimes\sigma_y\otimes\sigma_z)\    
\Ru^{t_1t_2}(\t_1,\t_2,\t_3)\ (\sigma_y\otimes\sigma_y\otimes\sigma_z),
\\[.5cm]  
&P_{13}\ \Ru^{t_1t_2t_3}\ (\t_3,\t_2,\t_1)\ P_{13}& =& 
(\tau\otimes\tau\otimes\tau)\ \Ru(\t_1,\t_2,\t_3)\ 
(\tau\otimes\tau\otimes\tau)^{-1}\,,
\end{array} 
\end{equation}
where the
superscripts $t_i$, \ $i=1,2,3$, denote the matrix transposition in the
$i$-th space, $P_{13}$ denotes the operator permuting the spaces $1$
and $3$ and 
$\sigma_x$, $\sigma_y$, $\sigma_z$ denote the Pauli matrices, 
\begin{equation}
\tau=\left(\begin{array}{cc} 1 & 0 \\ 0 &
  \ii\end{array}\right)\;,\qquad \tau^2=\sigma_z\,.
\end{equation}
In the following the symmetry properties will not play any
essential role, since the elliptic parameterisation
considered below explicitly breaks the spatial symmetry.

\begin{figure}[hbt]
\centering
\includegraphics[height=9cm]{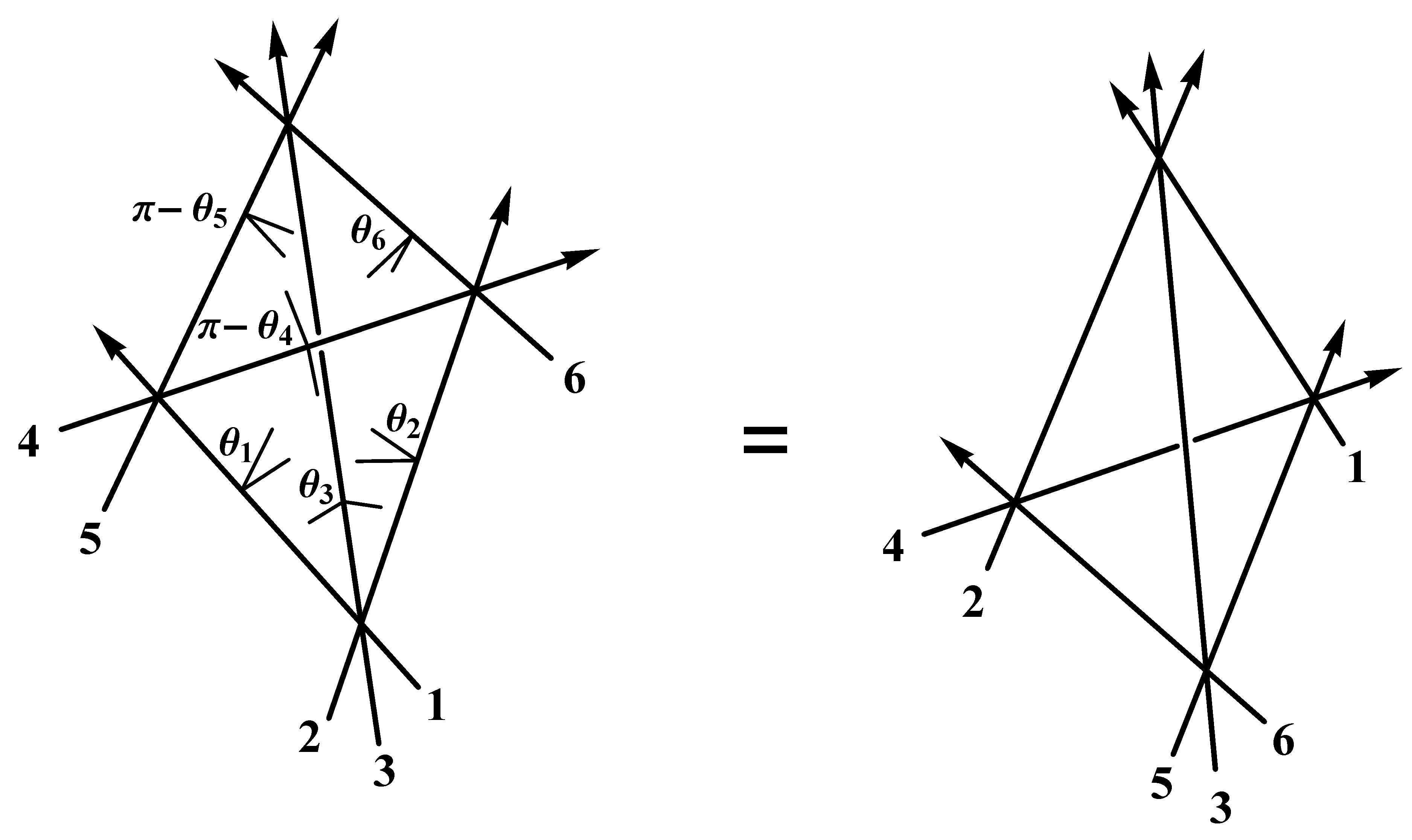}
\caption{Graphical representation of the tetrahedron equation
  \eqref{TE} showing the arrangement of the {\em internal} dihedral
  angles of the 
  tetrahedron 
}\label{fig2}
\end{figure}

The operator $\Ru$ satisfies the
tetrahedron equation,   
\beq
\begin{array}{l}
\Ru_{123}(\t_1,\t_2,\t_3)\;\Ru_{145}(\t_1,\t_4,\t_5)\;
\Ru_{246}(\pi-\t_2,\t_4,\t_6)\;\Ru_{356}(\t_3,\pi-\t_5,\t_6)\\[.5cm]  
\ \ \ \ =\Ru_{356}(\t_3,\pi-\t_5,\t_6)\;\Ru_{246}(\pi-\t_2,\t_4,\t_6)\;
\Ru_{145}(\t_1,\t_4,\t_5)\;\Ru_{123}(\t_1,\t_2,\t_3) 
,.\label{TE}
\end{array}
\eeq
which ensures the integrability of the model. The above equation
involves operators acting in a direct product ${\mathcal
  V}=(\C)^{\otimes\,6}$, of six identical two-dimensional vector
spaces\footnote{%
We follow usual notations, where 
$R_{123}$ acts non-trivially in
the first three spaces and coincides with the identity operator in all
the remaining components of the product. The other operators in
\eqref{TE}  are defined similarly.}. 
Geometrically, it is
associated with an arbitrary intersection of four planes. 
The indices in \eqref{TE} numerate six
intersection lines among these planes, which form extended edges
of an Euclidean tetrahedron. Each of the operators
$\Ru_{ijk}$ depends on its own set of spectral parameters, determined
by the dihedral angles corresponding to the lines $i$, $j$ and
$k$.
The geometric arrangement of these angles is shown in Fig.\ref{fig2}
(note that it is exactly the same as in \cite{Sergeev:1995rt}, but
different from that used in Eq. (2.2) in \cite{Baxter:1983qc}).
Altogether
there are six dihedral angles corresponding to six edges of the tetrahedron. 
At this point it is worth mentioning that these angles are not
independent; they satisfy one
non-linear constraint,
\beq
\det
\left(
\begin{array}{cccc}
1&{\phantom{+}}\cos \t_1&{\phantom{+}}\cos \t_2&\ -\cos \t_4\\
{\phantom{+}}\cos \t_1& 1 &{\phantom{+}}\cos \t_3&-\cos \t_5\\
{\phantom{+}}\cos \t_2&{\phantom{+}}\cos \t_3&1&{\phantom{+}}\cos\t_6\\
-\cos\t_4 &-\cos\t_5&{\phantom{+}}\cos\t_6&1 
\end{array}
\right)=0\,,
\eeq
which follows from the vanishing of the Gram determinant of four
unit normal vectors to faces of a tetrahedron in the Euclidean 3-space.
Therefore, Eq.\eqref{TE} contains only five
independent parameters. In matrix form it reads 
\beq
\begin{array}{l}
{\ds\sum_{{k_1\,k_2\,k_3,}\atop {k_4\,k_5\,k_6}}}
R(\t_1,\t_2,\t_3)^{k_1 k_2 k_3}_{i_1\, i_2\, i_3}\ 
R(\t_1,\t_4,\t_5)^{j_1\,k_4k_5}_{k_1i_4\,i_5}\ 
R(\pi-\t_2,\t_4,\t_6)^{j_2j_4k_6}_{k_2k_4i_6}\ 
R(\t_3,\pi-\t_5,\t_6)^{j_3\,j_5\,j_6}_{k_3k_5k_6}\\[.8cm]
={\ds\sum_{k_1,k_2,k_3,\atop k_4,k_5,k_6}}
R(\t_3,\pi-\t_5,\t_6)^{k_3k_5k_6}_{i_3\,i_5\,i_6}\ 
R(\pi-\t_2,\t_4,\t_6)^{k_2k_4j_6}_{i_2\,i_4\,k_6}\ 
R(\t_1,\t_4,\t_5)^{k_1j_4\,j_5}_{i_1\,k_4k_5}\ 
R(\t_1,\t_2,\t_3)^{j_1\,j_2\,j_3}_{k_1k_2k_3}\,.
\end{array}\label{TE-mat}
\eeq
Altogether there are $2^{12}$ distinct equations, half of
which is non-trivial (for the other half both sides vanish identically  
due to the parity conservation \eqref{parity}). 
There are several proofs that the weights \eqref{weights} indeed
satisfy these equations. Firstly, thanks to the ``cube-vertex''
correspondence of ref.\cite{Sergeev:1995fb}, this follows from 
the Baxter's proof \cite{Baxter:1983qc} of the tetrahedron equation
for the interaction-round-a-cube (IRC) formulation of the
model. Baxter used various symmetry relations and a computer-aided
analysis of a pattern of signs in some equations to reduce all
non-trivial equations
to just two equations, which were then proven with the help of
spherical trigonometry. Secondly, there exists 
a completely algebraic proof of \eqref{TE-mat}
given in \cite{Sergeev:1995rt}. This proof (as well as its earlier
variant \cite{KMS:1993} for the IRC formulation) is based on a
repeated application of the so-called {\em restricted star-triangle
  relation}, invented in \cite{Bazhanov:1993j}. It is worth
mentioning, that this relation was later interpreted as the five-term
identity for a cyclic version \cite{Bazhanov:1995jpa} 
of the quantum dilogarithm \cite{Faddeev:1993rs}. Finally, an
alternative (and, in fact, simpler) algebraic proof of \eqref{TE-mat},
based on the auxiliary linear problem, was obtained in \cite{Sergeev:1999jpa}.

To conclude this Section, 
mention a simple fact that Eq.\eqref{TE} is unaffected by
arbitrary linear similarity transformations in any of the six vector
spaces therein,
\beq
\Ru_{ijk}\to (\Gu_i\otimes\Gu_j\otimes\Gu_k)\,\Ru_{ijk}\,
(\Gu_i\otimes\Gu_j\otimes\Gu_k)^{-1}\,,\label{simil}
\eeq
where $\Gu_i$, $i=1,2,\ldots,6$ are arbitrary non-degenerate matrices.
We will use this freedom to bring the expressions for matrix elements
of $\Ru(\t_1,\t_2,\t_3)$ to the most convenient form.

\section{Elliptic parameterisation of the weights}
\subsection {Linear transformations of the weights} 
For the following analysis it is convenient to introduce the operator 
\beq
\L_{123}=\left({\bf
  D(\xi)}\otimes\Fu\otimes\Fu\right)\  \Ru_{123}\ \left({{\bf 
    D}(\xi)}\otimes
{\Fu}\otimes  {\Fu}\right)^{-1},\label{L-def}
\eeq
where
\beq
\Fu=\frac{1}{\sqrt{2}}\left(\begin{array}{lr}
  1&1\\1&-1 \end{array}\right) \ , \quad
{\bf D}(\xi)=\left(\begin{array}{lr}
  1&0\\0&\xi \end{array}\right) \ , \label{F-def}
\eeq
which differs from $\Ru_{123}$ by a simple similarity transformation
of the type \eqref{simil}.
The parameter $\xi$ will be specified later on. Evidently, the
transformation breaks the symmetry between the three spaces in
\eqref{L-def}, since the first space is distinguished from the other two.
The matrix elements
of $\Lu_{123}$ can be arranged into a set of $4\times4$  matrices
$L_0^0$, $L_0^1$, 
$L_1^0$ and $L_1^1$, acting in the tensor product of the second and
third spaces, while the indices $0,1$, labeling these matrices, refer to the
first space,
\begin{equation}
\begin{array}{rclrcl}
(L_0^0)_{2,3}^{}&=&\left(\begin{array}{cccc}
a & 0 & 0 & d\\
0 & b & c & 0\\
0 & c & b & 0\\
d & 0 & 0 & a
\end{array}\right)\;,\quad
&(L_1^1)_{2,3}^{}&=&\left(\begin{array}{cccc}
b & 0 & 0 & -c\\
0 & a & -d & 0\\
0 & -d & a & 0\\
-c & 0 & 0 & b
\end{array}\right)\;,
\end{array}\label{set1}
\end{equation}
where 
\begin{equation}
\begin{array}{rclrcl}
a&=&1-t_0t_1+t_2t_3+t_0t_1t_2t_3,\qquad & b&=&1+t_0t_1-t_2t_3 +
t_0t_1t_2t_3\;,\\[3mm]
c&=&1+t_0t_1+t_2t_3-t_0t_1t_2t_3\qquad & d&=&1-t_0t_1-t_2t_3-t_0t_1t_2t_3\;.
\end{array}\label{var1}
\end{equation}
and
\begin{equation}
\begin{array}{rclrcl}
(L_0^1)_{2,3}^{}&=&\xi^{-1}\left(\begin{array}{cccc}
-a' & 0 & 0 & -d'\\
0 & -b' & -c' & 0\\
0 & c' & b' & 0\\
d' & 0 & 0 & a'
\end{array}\right)\;,\quad
&(L_1^0)_{2,3}^{}&=&\xi\,\left(\begin{array}{cccc}
-b' & 0 & 0 & c'\\
0 & -a' & d' & 0\\
0 & -d' & a' & 0\\
-c' & 0 & 0 & b'
\end{array}\right)\;,
\end{array}\label{set2}
\end{equation}
where
\begin{equation}
\begin{array}{rclrcl}
a'&=&-t_1t_2-t_0t_3 +i t_0t_2 +i t_1t_3\;, \qquad& b'&
=&-t_1t_2-t_0t_3-i t_0t_2 - i t_1t_3\;,\\[3mm]
c'&=&-t_1t_2+t_0t_3+i t_0t_2 -i  t_1t_3\;,\qquad & d'&=&-t_1t_2+t_0t_3
- i t_0t_2 + i t_1t_3\;. 
\end{array}\label{var2}
\end{equation}
We regard these matrices as two by two block matrices with
2-dimensional blocks. The indices $(i_2,j_2)$ numerate the blocks
while the indices $(i_3,j_3)$ numerate matrix elements inside the
blocks. From \eqref{const} it follows that 
\beq ab =a' b',\qquad c d =c' d',\qquad 
a^2+b^2-c^2-d^2=0,\qquad a'^2+b'^2-c'^2-d'^2=0,\label{abcd-cond}
\eeq
therefore each of the above matrices can be viewed as an
$R$-matrix of the eight-vertex free-fermion model, which satisfy the condition
\beq
{\mathcal R}^{{\rm(FFM)}}=\left(\begin{array}{cccc}
\w_1 & 0 & 0 & \w_7\\
0 & \w_3 & \w_5 & 0\\
0 & \w_6 & \w_4 & 0\\
\w_8 & 0 & 0 & \w_2
\end{array}\right)\,,\qquad \w_1\w_2+\w_3\w_4-\w_5\w_6-\w_7\w_8=0\ .
\label{ffm}
\eeq

\subsection{Elliptic parameterisation of the weights} 
\begin{figure}[hbt]
\centering
\includegraphics[height=9cm]{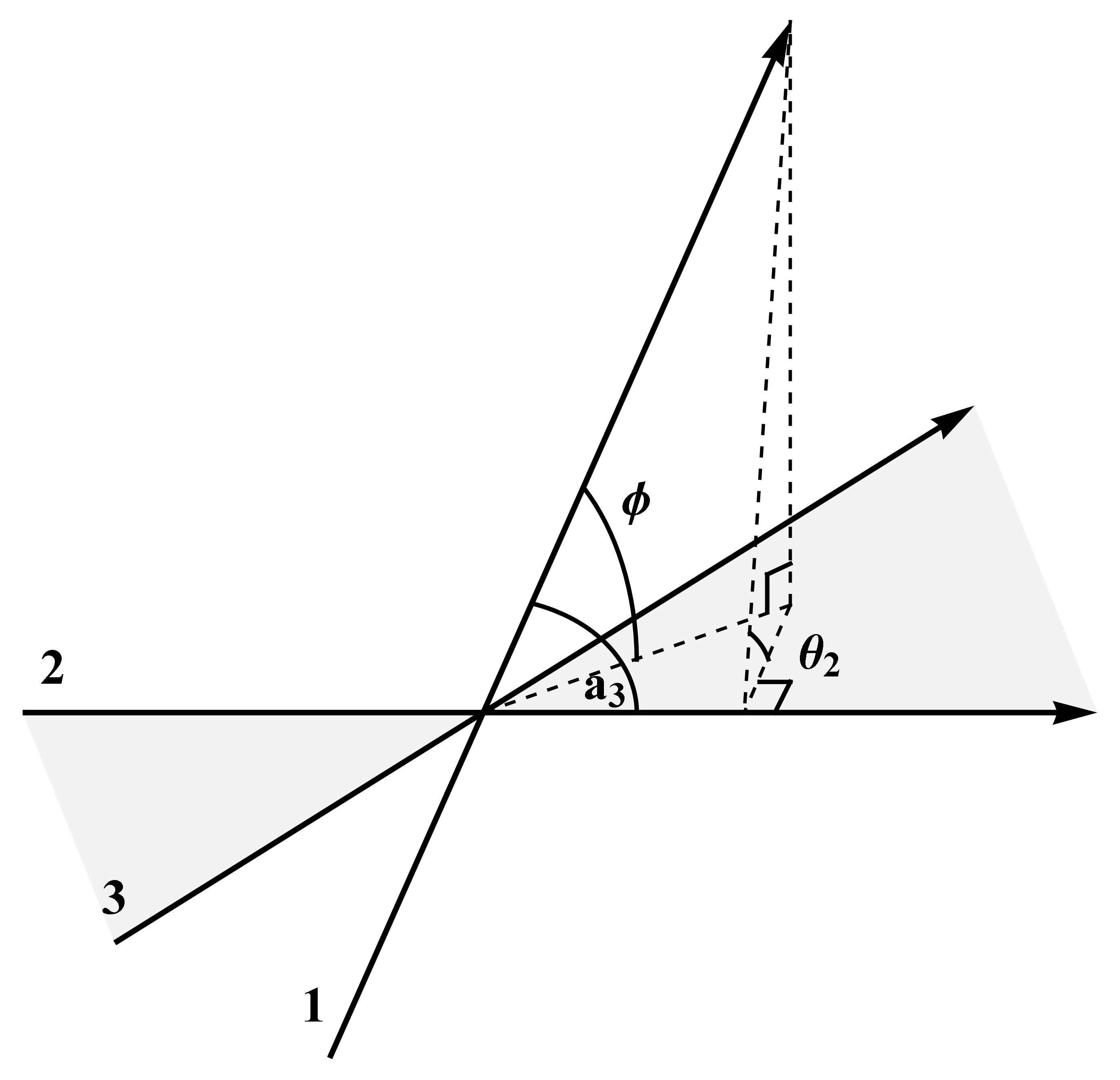}
\caption{This figure clarifies the geometric interpretation for the angle
  $\phi$, given by \eqref{fi-def}, as an angle between the line 1 and
  the plane containing the lines 2 and 3. 
}\label{fig3}
\end{figure}

Here we will show that the weights \eqref{weights} can be parametrised
in terms of Jacobi elliptic 
functions. First, note that the form of the matrices \eqref{set1} and
\eqref{set2}  
and the relations \eqref{abcd-cond} suggest to use Baxter's
parameterisation of 
the eight-vertex model 
\cite{Bax72} (the latter contains the symmetric free-fermion model
as a particular case). To do this define the elliptic modulus 
\beq
k=\frac{cd}{ab}\equiv\frac{c'd'}{a'b'}\label{modulus}\,,
\eeq
similarly to that in the symmetric eight-vertex model with the weights
$a,b,c,d$. (cf. Eq.(5.7) of
\cite{Bax72}). With an account of \eqref{var1} and \eqref{var2} 
the RHS of Eq.\eqref{modulus} is a function of the dihedral angles 
$\t_1,\t_2,\t_3$. After elementary simplifications, one obtains
\beq
k=\frac{1-\sin \phi}{1+\sin \phi},\qquad
\sin \phi=\frac{4t_0t_1t_2t_3}{(1-t_0^2t_1^2)(1-t_2^2t_3^2)}
=2\frac{\sqrt{\sin \alpha_0\sin\alpha_1\sin\alpha_2\sin\alpha_3}}{\sin\theta_1}
\;.\label{k-def}
\eeq
Then using the relations (see \S132 of \cite{Todhunter:1914}) 
\begin{equation}
\sin\alpha_0=2K\sin\frac{a_1}{2}\sin\frac{a_2}{2}\sin\frac{a_3}{2}\;,\quad 
\sin\alpha_i=2K\sin\frac{a_i}{2}\cos\frac{a_j}{2}\cos\frac{a_k}{2}\;,
\end{equation}
where $(i,j,k)$ is any permutation
of $(1,2,3)$  and $K$ is defined in \eqref{sine}, 
one obtains  
\begin{equation}
\sin\phi =K^2\frac{\sin a_1}{\sin \theta_1}\sin a_2 \sin a_3 = \sin
\t_2\sin a_3\label{fi-def}
\end{equation} 
which shows that $\phi$ 
is the angle between the line $1$ and 
the plane containing the lines $2$ and $3$, as illustrated in Fig.\ref{fig3}. 
Note that value $k=0$ corresponds to the right angle
$\phi=\pi/2$. 

Further, let $\sn\, x=\sn(x,k)$, $\cn\, x=\cn(x,k)$, $\dn\, x 
=\dn(x,k)$ and $\cd \,x=\cd(x,k)$ 
denote Jacobi elliptic functions of the argument $x$ and modulus $k$
(we follow the 
notations of \cite{WW}). Now let us parametrise the angles $\t_2$ and
$\t_3$ in terms of new parameters $w_1$ and $w_2$,
\beq
e^{-i\t_2}=-\frac{(1+\sn(2w_1))(1-k\,\sn(2w_1))}{\cn(2w_1)
\dn(2w_1
 )},\qquad
e^{i\t_3}=\frac{(1+\sn(2w_2))(1-k\,\sn(2w_2))}{\cn(2w_2)\dn(2w_2)}.
\label{u12-def}
\eeq
Given $\t_2,\t_3$ each of these equations have two solutions in the
periodicity rectangle (we assume $0< k< 1$)
\beq
0\le w_1,w_2 < 2 {\bf K},\qquad 0\le
\mathop{{\rm Im}}w_1,\mbox{Im}\,w_2< i {\bf K}'/2\ .
\eeq
where ${\bf K}={\bf K}(k)$ and ${\bf K}'={\bf K}(\sqrt{1-k^2})$
are   the complete
elliptic integral of the first kind. Now substitute \eqref{u12-def} back
into \eqref{k-def} and solve the resulting equation 
for $\t_1$ in terms $k, w_1,
w_2$. There are four solutions of the form 
\beq
e^{i\t_1}=\frac{\cd(2w_2)}{\cd(2w_1)}\,,\label{t1u}
\eeq
corresponding to four possible choices of $w_1,w_2$ in
\eqref{u12-def}. Only one of them\footnote{%
Calculating $\t_1$ from \eqref{t1u} for all four solutions of
\eqref{u12-def} one
obtains the following values $\t_1=\pm \t_1^{(true)}, \pm \t_1'$\,
where $\t_1^{(true)}$ is the original value of $\t_1$, while $\t_1'$
is a completely different value, 
corresponding to a different spherical triangle,
$(\t_1',\t_2,\t_3)$ which, nevertheless, 
leads to the same modulus $k$ in equation \eqref{k-def}.} 
lead to the original value of $\t_1$. This  allows one 
 to uniquely fix the required solution of \eqref{u12-def}.

From now on we regard $k, w_1, w_2$
as new independent variables instead of $\t_1,\t_2,\t_3$. The latter
are now defined by the formulae $\eqref{u12-def}$ and $\eqref{t1u}$.
The weights in Eq.\eqref{weights} are expressed through the square roots
of tangents of halves of spherical excesses
$\alpha_0,\alpha_1,\alpha_2,\alpha_3$. Surprisingly enough,  after 
lengthy calculations we found that these
tangents have rather simple expressions in the new variables,
\beq
\begin{array}{rclrcl}
t_0&=&\Big({{-if({w}_-)}\big/{f({\bf K}-{w}_+)}}\Big)^{\frac{1}{2}},\qquad
&t_1&=&\Big({if({w}_-)\,f({\bf K}-{w}_+)}\Big)^{\frac{1}{2}},\\[.3cm]
t_2&=&\Big({-if({\bf K}-{w}_-)\,f({w}_+)}\Big)^{\frac{1}{2}},\qquad
&t_3&=&\Big({if({\bf K}-{w}_-)\big/f({w}_+)}\Big)^{\frac{1}{2}},
\end{array}
\eeq
where 
\beq
{w}_\pm={w}_1\pm{w}_2,\qquad k'=\sqrt{1-k^2}\ .
\eeq 
The function $f(w)$ is defined as  
\beq
f({w})=\frac{k'^2\,\sn {w}}{(\cn {w}+\dn{w})(k\,\cn{w}+\dn{w})}\ .
\eeq
It follows that the products appearing in \eqref{var1} and \eqref{var2}
simplify to 
\beq
\begin{array}{rclrcl}
t_0t_1&=&f({w}_-)\,,\quad& t_2t_3&=&f(K-{w}_-)\,,\\[.3cm]
t_0t_3&=&\ds\left({\frac{f({w}_-)\,f({\bf
      K}-{w}_-)}{f({w}_+)\,f({\bf K}-{w}_+)}}\right)^{\frac{1}{2}}, 
\quad
&t_1t_2&=&f({w}_+)\,f({\bf K}-{w}_+)t_0t_3\ .
\end{array}
\eeq
Then using addition theorems for elliptic functions, one obtains
\beq
\begin{array}{llll}
a=\rho_-\,\cd{w}_-\,,\quad
&b=\rho_-\,\sn{w}_-\,,\quad &c=\rho_-\,,\quad& d=\rho_-\, k\,
\cd{w}_-\,\sn{w}_-\,,\\[.3cm] 
a'=\rho_+\,\cd{w}_+\,,\quad&
b'=\rho_+\,\sn{w}_+\,,\quad& c'=\rho_+\,,\quad& d'=\rho_+\, k\,
\cd{w}_+\,\sn{w}_+\,, 
\end{array}
\eeq
where 
\beq
\begin{array}{rclcl}
\rho_-&=&1+t_0t_1+t_2t_3-t_0t_1t_2t_3&=&\ds\frac{4(1-\sn{w}_-)\dn{w}_-}
{\cn{w}_-\,\left(\cn{w}_-\,\dn{w}_-+(1-\sn{w}_-)(1+k\,\sn{w}_-)\right)}\,,
\\[.5cm]
\rho_+&=&t_0t_3-t_1t_2+i(t_0t_2-t_1t_3)
&=&-\rho_-\ \left(\ds\frac{\cd{w}_-\,\sn{w}_-}{\cd{w}_+\,\sn{w}_+}\right)^{\frac{1}{2}}\ .
\end{array}\label{rhopm}
\eeq
Next, we specify the parameter $\xi$ in \eqref{L-def}, which so far
remained at our disposal,  
\beq
\xi=-\frac{\rho_+}{\rho_-}\equiv h(w_1,w_2),\qquad
h(x,y)=\sqrt{\frac{\cd(x-y)\,\sn(x-y)}{\cd(x+y)\,\sn(x+y)}}.\label{xidef}
\eeq
Finally, introduce 
the $R$-matrix of the symmetric eight-vertex model \cite{Bax72}
specialized to the free-fermion case,  
\beq
{\mathbb R}(w)=
\left(\begin{array}{cccc}
\cd w & 0 & 0 & k\,\cd w \,\sn w \\
0 & \sn w & 1 & 0\\
0 & 1 & \sn w & 0\\
k\,\cd w \,\sn w & 0 & 0 & \cd w
\end{array}\right)\;\label{Rffm}
\eeq
Using this notation, the matrices \eqref{set1}, \eqref{set2} can be
written in a uniform way 
\beq
\begin{array}{rclrcl}
(L_0^0)_{23}&=&\phantom{i\,}\rho_-\, {\mathbb R}_{23}(w_-),\qquad 
&(L_0^1)_{23}&=&-\rho_+\,\xi^{-1}\,\sigma_z^{(2)}\,{\mathbb
    R}_{23}(w_+),\\[.3cm] 
(L_1^0)_{23}&=&i\,\ds{\rho_+}\,\xi\;\sigma_x^{(2)}\,{\mathbb
    R}_{23}(w_+)\,\sigma_y^{(2)},\qquad
&(L_1^1)_{23}&=&\phantom{-}\rho_-\, \sigma_y^{(2)}\,{\mathbb
    R}_{23}(w_-)\,\sigma_y^{(2)},  
\end{array}\label{uniform}
\eeq
where the Pauli matrices
$\sigma_x^{(2)},\sigma_y^{(2)},\sigma_z^{(2)}$ act in the space $2$.
Taking into account \eqref{rhopm} and \eqref{xidef} is easy to see that all
expressions for the matrix elements in \eqref{uniform} are {\em
  meromorphic double periodic functions} of the variables $w_1$ and $w_2$.

To summarize, we have shown that all matrix elements the operator
\beq
{\mathcal L}_{123}(w_1,w_2\,|\,k)=\Lu_{123}(\t_1,\t_2,\t_3)\bedef
\left({\bf
  D(\xi)}\otimes\Fu\otimes\Fu\right)\  \Ru_{123}(\t_1,\t_2,\t_3)\ \left({{\bf 
    D}(\xi)}\otimes
{\Fu}\otimes  {\Fu}\right)^{-1},\label{Lnew}
\eeq
are meromorphic functions of $w_1$ and $w_2$, provided the two sets
of variables, entering different sides of this equation, are related by
\eqref{k-def}, \eqref{u12-def} and \eqref{t1u} and the parameter $\xi$
is defined by \eqref{xidef}.

We would like to stress the above elliptic parametrisation was
obtained for a generic case without any
special requirements for the values of $\t_1,\t_2,\t_3$.   
By construction, this parametrisation breaks the
symmetry of the weights 
(the directions $1$ is distinguished). Moreover, the modulus $k$ 
will, {\em a priori},  be different for different vertices of 
the tetrahedron. Therefore, in general, one cannot  
parametrise all weights in the tetrahedron
equation by elliptic functions of the same modulus. Nonetheless, there
exists an important four-parameter reduction of this equation, where
such parametrisation is possible. This is the ``prismatic limit''
considered in the next section.

\section{Reductions of the tetrahedron equation}
In this section we consider certain reductions of the tetrahedron
equation for the solution \eqref{weights} 
and its connection to an example of the {\em tetrahedral
  Zamolodchikov algebra} constructed in \cite{Korepanov:1993}.  

\subsection{The ``static limit''}\label{sec:statlim}
The term ``static limit'' stems from the original Zamolodchikov's 
work \cite{Zamolodchikov:1980rus} where
it was related to the case of slowly moving or ``non-relativistic''
straight strings in $2+1$ dimensions. 
On the level of parameters this corresponds to a
configuration where the sum of dihedral angles 
is equal to $\pi$ for every vertex of the tetrahedron. Namely, for
Eq.\eqref{TE-mat} this means
\beq\begin{array}{rclrcl}
\t_1+\t_2+\t_3&=&\pi,\qquad& \t_1+\t_4+\t_5&=&\pi,\\[.3cm]
\t_4+\t_6-\t_2&=&0,\qquad &\t_3+\t_6-\t_5&=&0\ .\label{static}
\end{array}
\eeq
The linear
angles between edges at the vertices all become equal to $0$ or $\pi$ and
the tetrahedron degenerates into four planes intersecting along the
same line. Their relative orientation is fixed by three angles only, so the   
tetrahedron equation in this case contains three independent parameters
(instead of five in the general case).

We will use a special notation
for the operator $\Ru(\t_1,\t_2,\t_3)$ specialized to this case
\beq
{\bf
  S}(\t_1,\t_2,\t_3)=\Ru(\t_1,\t_2,\t_3)\,,\qquad\qquad  \t_1+\t_2+\t_3=\pi\ . 
\label{S-def}
\eeq 
Its matrix elements are determined by \eqref{weights} where one sets
\beq
t_0=0, \qquad t_i=T_i\equiv\sqrt{\tan\frac{\t_i}{2}},\qquad
i=1,2,3.\label{t0zero} 
\eeq
There are only sixteen non-vanishing elements,
\beq
\begin{array}{rcrcrcrcc}
S_{0,0,0}^{0,0,0}&= & S_{0,1,1}^{0,1,1}&= & S_{1,0,1}^{1,0,1}&= & S_{1,1,0}^{1,1,0}&= & 1\\ \\

S_{0,0,1}^{0,1,0}&= & S_{0,1,0}^{0,0,1}&= & -S_{1,1,1}^{1,0,0}&= & -S_{1,0,0}^{1,1,1}&= & \phantom{-\ii}\,T_2T_3\\ \\
S_{0,1,0}^{1,1,1}&= & S_{0,0,1}^{1,0,0}&= & -S_{1,0,0}^{0,0,1}&= & -S_{1,1,1}^{0,1,0}&= & -\ii\, T_1T_3\\ \\
S_{1,1,1}^{0,0,1}&= & S_{0,0,1}^{1,1,1}&= & S_{0,1,0}^{1,0,0}&= & S_{1,0,0}^{0,1,0}&= & \phantom{-\ii}\,T_1T_2\,.
\end{array}\label{weights2}
\eeq
This solution of the tetrahedron equation (with a slightly different
parameterisation, see Sect.\ref{sec:tza} below) was first obtained in
\cite{Korepanov:1993}. Two years 
later 
\cite{Sergeev:1995rt} it was understood as the vertex form of the
original Zamolodchikov's solution in the static limit
\cite{Zamolodchikov:1980rus}.

The operator ${\bf S}$ satisfies the relation 
\beq
{\bf S}^2=1.\label{sqone}
\eeq
It has a block-diagonal form with two $4\times4$ blocks, a trivial
one, which coincides with the identity matrix, and a non-trivial one which
has two eigenvalues $+1$ and two eigenvalues $-1$.

The operator ${\bf S}$ possesses left and right ``bare vacuum'' eigenvectors 
\beq
({v_0}\otimes{v_0}\otimes{v_0})^t\,{\bf S}=({v_0}\otimes{v_0}\otimes{v_0})^t\,,
\quad{\bf S}\,({v_0}\otimes{v_0}\otimes{v_0})=
{v_0}\otimes{v_0}\otimes{v_0}\,,\qquad v_0=\left({ 1
  \atop 0} \right).\label{v0-def}
\eeq
Note that this property does not hold for the general case
\eqref{weights} with $t_0\not=0$.
\subsection{The planar limit}
Geometrically the ``planar limit'' corresponds to the case when all
four vertices of the tetrahedron lie in one plane, i.e. when the
tetrahedron becomes ``squashed'' into a plane. An edge-spin solution of
\eqref{TE} corresponding to this case was first obtained by Hietarinta
\cite{Hietarinta:1994}.
Subsequently it was understood as the planar limit
\cite{Mangazeev:1994ut} 
of the
Zamolodchikov model. The details of this reduction are rather
complicated and not immediately related to the main topic of this
paper. Therefore, we refer interested readers to the original publication
\cite{Mangazeev:1994ut} where this limit is thoroughly studied. 
\subsection{The prismatic limit}
Here we consider yet another limiting case of the equation \eqref{TE},
when one of the
tetrahedron vertices goes to infinity (we choose it to be the vertex
corresponding to $\Ru_{123}$). Then the sum of dihedral angles
at this vertex will satisfy an additional constraint, 
\beq
\t_1+\t_2+\t_3=\pi,\label{t123}
\eeq
whereas the whole tetrahedron turns into an infinite
prism. The number
of independent angles reduces from five to four. The edges $1$,
$2$ and $3$ become parallel and therefore, 
have the same angle to the plane, containing the edges $4,5$ and $6$,
forming  the base of the prism. Remembering the geometric definition
\eqref{k-def} of the elliptic modulus $k$ in the previous section, we
conclude that all weights corresponding to the vertices $(1,4,5)$ ,
$(2,4,6)$ and $(3,5,6)$ can be
parametrised by elliptic functions of the {\em same modulus}.

Eqs. \eqref{k-def}, \eqref{u12-def} and \eqref{t1u} 
define a generic change of variables
from three dihedral angles $(\t_1,\t_2,\t_3)$ to new parameters
$(k,w_1,w_2)$. Here we want to apply this
substitution for three different sets of dihedral angles
corresponding to the three vertices at the base of the prism
\beq
(\t_1,\t_4,\t_5)\to (k,u_1,u_2);\quad 
(\pi-\t_2,\t_4,\t_6)\to (k,u_1,u_3);\quad 
(\t_3,\pi-\t_5,\t_6)\to (k,u_2,u_3).\quad\label{three} 
\eeq
As explained above from the geometric considerations the elliptic
modulus $k$ will 
automatically be the same for  all three sets. For instance, let
$(b_1,b_2,b_3)$ be the sides of the spherical triangle with angles 
$(\t_1,\t_4,\t_5)$, then we set  
\beq
k=\frac{1-\sin \phi}{1+\sin \phi},\qquad \sin\phi=\sin\t_4 \,\sin b_3\,.
\eeq
Next define the variables $u_1,u_2,u_3$ by the relations
\bea
&&e^{-i\t_4}=-\frac{(1+\sn(2{u}_1))(1-k\,\sn(2{u}_1))}{\cn(2{u}_1)\dn(2{u}_1)},\nonumber\\
&&e^{i\t_5}\phantom{n}=\phantom{-}\frac{(1+\sn(2{u}_2))(1-k\,\sn(2{u}_2))}{\cn(2{u}_2)\dn(2{u}_2)},\label{teta456}\\
&&e^{i\t_6}\phantom{n}=\phantom{-}\frac{(1+\sn(2{u}_3))(1-k\,\sn(2{u}_3))}{\cn(2{u}_3)\dn(2{u}_3)}\nonumber
\eea
and 
\beq
e^{i\t_1}=\frac{\cd(2{u}_2)}{\cd(2{u}_1)},\quad
e^{i\t_2}=-\frac{\cd(2{u}_1)}{\cd(2{u}_3)},\quad
e^{i\t_3}=\frac{\cd(2{u}_3)}{\cd(2{u}_2)}\label{teta123}
\eeq
which are obtained by a simple specialisation of \eqref{u12-def} and
\eqref{t1u} for the three substitutions \eqref{three}. Note that the above
formulae \eqref{teta456} and \eqref{teta123} give a parameterisation 
of six angles of a triangular prism in terms of elliptic functions. 
In particular, it is easy to check that the 
three expressions in \eqref{teta123} are consistent with \eqref{t123}.

Let us now rewrite the tetrahedron equation \eqref{TE} 
for this case in the new variables. We will do this in several steps.
First, guided by the formula \eqref{Lnew} define three operators 
\beq\begin{array}{rcl}
{\mathcal L}_{145}(u_1,u_2)&=&
\left({\bf
  D}(\xi_1)\otimes\Fu\otimes\Fu\right)\  \Ru_{145}(\t_1,\t_4,\t_5)\ 
\left({{\bf D}(\xi_1)}\otimes
{\Fu}\otimes  {\Fu}\right)^{-1},\\[.3cm]
{\mathcal L}_{246}(u_1,u_3)&=&
\left({\bf
  D}(\xi_2)\otimes\Fu\otimes\Fu\right)\  
\Ru_{246}(\pi-\t_2,\t_4,\t_6)\ \left({{\bf 
    D}(\xi_2)}\otimes
{\Fu}\otimes  {\Fu}\right)^{-1},\\[.3cm]
{\mathcal L}_{356}(u_2,u_3)&=&
\left({\bf
  D}(\xi_3)\otimes\Fu\otimes\Fu\right)\  \Ru_{145}(\t_3,\pi-\t_5,\t_6)\ 
\left({{\bf D}(\xi_3)}\otimes
{\Fu}\otimes  {\Fu}\right)^{-1},
\end{array}\label{Lthree}
\eeq
Here ${\bf F}$ and ${\bf D}(\xi)$ are defined in \eqref{F-def} and the
parameters $\xi_1,\xi_2,\xi_3$ have been chosen in agreement with
\eqref{xidef} in each of the three cases, 
\beq
\quad \xi_1=h(u_1,u_2),\quad \xi_2=h(u_1,u_3),\quad
\xi_3=h(u_2,u_3),\label{xi123} 
\eeq
where $h(x,y)$ is defined in \eqref{xidef}. Therefore, according to
the result of Sect.3, all operators \eqref{Lthree} are
meromorphic double periodic functions of $u_1, u_2, u_3$, which, of
course, implicitly  
depend on the elliptic modulus $k$.

Next, we need to express in the new variables the weights, 
corresponding to the infinitely distant vertex $(1,2,3)$. 
It is convenient to define a new operator,
\beq
{\mathcal S}_{123}(u_1,u_2,u_3)=({\bf D}(\xi_1)\otimes{\bf D}(\xi_2)\otimes{\bf
  D}(\xi_3))\;\Su_{123}(\t_1,\t_2,\t_3)\;({\bf D}(\xi_1)\otimes{\bf
  D}(\xi_2)
\otimes{\bf
  D}(\xi_3))^{-1}\,,
\label{S2-def}
\eeq
which differs from \eqref{S-def} by a diagonal similarity transformation.

Using 
\eqref{teta123} it is not difficult to show that 
\beq
\tan\frac{\t_1}{2}=-ig(u_1,u_2),
\quad \tan\frac{\t_2}{2}=\frac{i}{g(u_1,u_3)},\quad
\tan\frac{\t_3}{2}=-ig(u_2,u_3),\label{tan123}
\eeq
where 
\beq
g(x,y)=k'^2\,
\frac{\sd(x-y)\sd(x+y)}{\cn(x-y)\cn(x+y)},\label{g-def}
\eeq
Combining this with \eqref{weights2}, 
one obtains the following expressions for the
non-vanishing matrix 
elements of $\Sc(u_1,u_2,u_3)$ 
\beq
\begin{array}{rcccccccl}
\Sc_{000}^{000}&=&\Sc_{110}^{110}&=&\Sc_{101}^{101}&=&\Sc_{011}^{011}&=&1,
\end{array}\label{Snew}
\eeq
$$
\begin{array}{rclrclrcl}
\Sc_{010}^{100}&=&U(u_1,u_2,u_3),\quad &\Sc_{010}^{001}&=&U(u_3,u_2,u_1),
\quad &S_{100}^{001}&=&V(u_1,u_2,u_3), \\[.3cm]
\Sc_{100}^{010}&=&U(u_1,-u_2,-u_3),\quad
&\Sc_{001}^{010}&=&U(u_3,-u_2,-u_1),\quad
&\Sc_{001}^{100}&=&V(-u_1,u_2,-u_3), \\[.3cm]
\Sc_{111}^{001}&=&U(u_1,-u_2,u_3),\quad
&\Sc_{111}^{100}&=&U(u_3,-u_2,u_1),\quad
&\Sc_{111}^{010}&=&V(u_1,u_2,-u_3),\\[.3cm]
\Sc_{001}^{111}&=&U(u_1,u_2,-u_3),\quad &\Sc_{100}^{111}&=&U(u_3,u_2,-u_1),
\quad &\Sc_{010}^{111}&=&V(-u_1,u_2,u_3),
\end{array}
$$
where
\beq
U(x,y,z)=\frac{\sd(x+y)\cn(x-z)}{\cn(x-y)\sd(x+z)},
\quad V(x,y,z)=-k'^2\,\frac{\sd(y-x)\sd(y+z)}{\cn(y+x)\cn(y-z)},
\eeq
It is easy to see that all above matrix elements are meromorphic
functions of $u_1, 
u_2, u_3$, implicitly depending on the elliptic modulus $k$.
Note that even though the parameters
$\xi_1,\xi_2,\xi_3$, entering  
\eqref{S2-def}, were fixed by
the analyticity requirements for the other three operators, defined in 
\eqref{Lthree}, the matrix elements of ${\cal
  S}(u_1,u_2,u_3)$ have automatically became  
meromorphic functions on the elliptic curve\footnote{%
Eq.\eqref{TE2} below implies that ratios of matrix elements of
${\mathcal S}_{123}(u_1,u_2,u_3)$ are meromorphic functions on the
elliptic curve. However, some of these matrix elements in \eqref{Snew} 
are equal to one, therefore the remaining elements should be meromorphic.}.   
Note also that the similarity transformation in
\eqref{S2-def} introduces two additional  
parameters into \eqref{Snew} and breaks the symmetry relations between
different 
matrix elements exhibited in \eqref{weights2}. Remind that the
original static limit weights \eqref{weights2} depend on only two
independent angles.  

To complete our analysis of the prismatic
limit apply the    
similarity transformation with the matrix 
\beq
{\bf G}={\bf D}(\xi_1)\otimes{\bf D}(\xi_2)\otimes{\bf
  D}(\xi_3)\otimes {\bf F}\otimes {\bf F}\otimes {\bf F}
\eeq
to both sides of the tetrahedron equation \eqref{TE}. Then using the 
definitions \eqref{Lthree} and \eqref{S2-def} one obtains
\beq\begin{array}{l}
\Sc_{123}(u_1,u_2,u_3)\,{\cal L}_{145}(u_1,u_2)\,{\cal
  L}_{246}(u_1,u_3)
\,{\cal L}_{356}(u_2,u_3)\\[.4cm]\qquad=\ 
{\cal L}_{356}(u_2,u_3)\,{\cal L}_{246}(u_1,u_3)\,{\cal
  L}_{145}(u_1,u_2)
\,{\cal S}_{123}(u_1,u_2,u_3)\ .
\end{array}
\label{TE2}
\eeq
As required, this equation contains exactly four independent
parameters, namely,
$u_1,u_2,u_3$, which are indicated explicitly, and the elliptic modulus $k$
which is implicitly assumed.

\subsection{Elliptic parameterisation of the static limit}

Consider the tetrahedron equation \eqref{TE} in the static limit,
i.e., when the angles satisfy the constraints \eqref{static}. In the
previous subsection we saw that upon the change of variables
\eqref{teta123} and the similarity transformation \eqref{S2-def} 
the vertex weights ${\cal S}_{123}(u_1,u_2,u_3)$, corresponding to the
vertex $(1,2,3)$, become meromorphic
functions on the elliptic curve. Remarkably, it is possible to do this
for the other three vertices as well, using the elliptic functions of
the same modulus. First, we need to parameterise the angles, satisfying
\eqref{static}, in term of the elliptic functions. To avoid confusions
with the notations of the previous subsection we denote these angles
$\ts_1,\ts_2,\ldots,\ts_6$. Let us parameterise $\ts_1,\ts_2,\ts_3$ by
the same formulae as \eqref{teta123},
\beq
e^{i\ts_1}=\frac{\cd(2{u}_2)}{\cd(2{u}_1)},\quad
e^{i\ts_2}=-\frac{\cd(2{u}_1)}{\cd(2{u}_3)},\quad
e^{i\ts_3}=\frac{\cd(2{u}_3)}{\cd(2{u}_2)}\label{teta123s}
\eeq
and assume that 
\beq
e^{i\ts_4}=-\frac{\cd(2{u}_1)}{\cd(2{u}_4)},\quad
e^{i\ts_5}=\frac{\cd(2{u}_4)}{\cd(2{u}_2)},\quad
e^{i\ts_6}=\frac{\cd(2{u}_4)}{\cd(2{u}_3)}\label{teta456-static}
\eeq
This implies the expressions \eqref{tan123} with $\t_1,\t_2,\t_3$ 
replaced by $\ts_1,\ts_2,\ts_3$ and 
\beq
\tan\frac{\ts_4}{2}=\frac{i}{g(u_1,u_4)},
\quad \tan\frac{\ts_5}{2}=-{i}{g(u_2,u_4)},\quad
\tan\frac{\ts_6}{2}=-ig(u_3,u_4),\label{tan456-static}
\eeq
where $g(u,w)$ is defined in \eqref{g-def}. Note, that these formulae
contains five independent parameters: $u_1,u_2,u_3,u_4$ and $k$,
whereas there are only three independent angles, satisfying
\eqref{static}. So the above parameterisation contains two spurious
parameters. 

Next, assume that $\xi_1,\xi_2,\xi_3$ are given by \eqref{xi123} and 
\beq
\quad \xi_4=h(u_1,u_4),\quad \xi_5=h(u_2,u_4),\quad
\xi_6=h(u_3,u_4),\label{xi456} 
\eeq
where $h(u,w)$ is defined in \eqref{xidef}. Applying now the
similarity transformation with the matrix 
\beq
{\bf \overline{G}}={\bf D}(\xi_1)\otimes{\bf D}(\xi_2)\otimes{\bf
  D}(\xi_3)\otimes {\bf D}(\xi_4)\otimes{\bf D}(\xi_5)\otimes{\bf
  D}(\xi_6)\label{Gbar}
\eeq
to both sides of the tetrahedron equation \eqref{TE} and taking into
account the above parameterisation, one obtains
\beq
\begin{array}{l}
\Sc_{123}(u_1,u_2,u_3)\,\Sc_{245}(u_1,u_2,u_4)\,
\Sc_{246}(u_1,u_3,u_4)\,\Sc_{356}(u_2,u_3,u_4)\\[.3cm]
\qquad=\ 
\Sc_{356}(u_2,u_3,u_4)\,\Sc_{246}(u_1,u_3,u_4)\,
\Sc_{245}(u_1,u_2,u_4)\,\Sc_{123}(u_1,u_2,u_3)
\end{array}\label{TE3}
\eeq
where the matrix elements of $\Sc_{ijk}(u,v,w)$ are defined by
\eqref{Snew}. It appears, that this equation 
naturally complements the equation \eqref{TE2}.
As noted in Sect.\ref{sec:statlim} there are only three
independent angles in the tetrahedron in the static limit. However,
Eq.\eqref{TE3} contains five independent parameters (four $u$'s and
the modulus $k$). Two additional parameters are not spurious in this
case, as they cannot be removed by a re-parameterisation. They were
introduced via the similarity transformation with the matrix \eqref{Gbar}.

\subsection{Tetrahedral Zamolodchikov algebra}\label{sec:tza}
Here we show that some additional reduction of the tetrahedron
equation \eqref{TE2} leads precisely to the example of the 
tetrahedral Zamolodchikov algebra constructed in \cite{Korepanov:1993}. 
It is convenient to 
introduce two $4\times4$ matrices, 
acting in the product of two spaces $\C\otimes\C$  as
\beq
{\cal R}^a(u,w)=\sum_{i_2,i_3,j_2,j_3}
 {\cal L}_{0\,i_2\,i_3}^{a\,j_2\,j_3}(u,w)\ 
(E_{i_2}^{j_2}\otimes E_{i_3}^{j_3}), \qquad a=0,1,\label{Ra-def}
\eeq
where $(E_i^j)_{ab}=\delta_{ia}\delta_{jb}$ denotes the $2\times2$ 
matrix unit.
According to \eqref{uniform}, \eqref{Lnew} 
these new matrices are
simply related to the $R$-matrix of the free-fermion model
\eqref{Rffm},
\beq
{\cal R}^{0}(u,w)=\rho_-\,{\mathbb R}(u-w),\qquad
{\cal R}^{1}(u,w)=-\rho_+\,\xi^{-1}\,(\sigma_z\otimes 1)\ {\mathbb
  R}(u+w)\,. 
\eeq
Now calculate matrix elements of tetrahedron equation 
\eqref{TE2} sandwiched between
the fixed vectors 
\beq
\langle L|=(v_0^t\otimes v_0^t\otimes v_0^t)_{123}\,,\qquad
| R\rangle=(v_a\otimes v_b\otimes v_c)_{123}
\eeq
in the first three spaces, where the basis vectors $v_0,v_1$ are
defined \eqref{v01-def} and the indices $a,b,c$ take the values
$0,1$ (in the notations of \eqref{TE-mat} this corresponds to the case
when $i_1=i_2=i_3=0$ and $j_1=a, j_2=b$, $j_3=c$). The
resulting equation considerably simplifies, since the matrix
$\Sc$ in the LHS of \eqref{TE2} drops out due to \eqref{v0-def}.
Writing this equation in an abbreviated
form, one gets
\beq\begin{array}{rcl}
({\cal L}_{0}^{a})_{45}\,({\cal
  L}_{0}^{b})_{46}
\,({\cal L}_{0}^{c})_{56}&= &\ds\sum_{d,e,f}\,
{\mathcal S}_{d\,e\,f}^{a\,b\,c}\,
({\cal L}_{0}^{f})_{56}\,({\cal L}_{0}^{e})_{46}\,({\cal
  L}_{0}^{d})_{45}
\end{array}
\label{ZTA0}
\eeq
where, for example 
\beq
({\mathcal L}_0^a)_{45}=
(v_0^t\otimes1\otimes1)\  {\mathcal L}_{145}\  (v_a\otimes
1\otimes1)
\eeq
and similarly for the other ${\cal L}'s$.
For fixed values of $a,b,c$, both sides of \eqref{ZTA0} still remain
operators acting in the spaces $4,5,6$. 
At this point it is convenient
to relabel these spaces as $1,2,3$ (the former spaces $1,2,3$ are no
longer required, so there should be no confusions) and use the
definition \eqref{Ra-def}. In this way one obtains  
\beq
\begin{array}{l}
{\cal R}^{a}_{12}(u_1,u_2)\,{\cal
  R}^{b}_{13}(u_1,u_3)
\,{\cal R}^{c}_{23}(u_2,u_3)
\\[.4cm]\qquad\qquad
=\ds\sum_{d,e,f}\,
{\mathcal S}_{d\,e\,f}^{a\,b\,c}(u_1,u_2,u_3)\,
{\cal R}^{f}_{23}(u_2,u_3)\,{\cal R}^{e}_{13}(u_1,u_3)\,{\cal
  R}^{d}_{12}(u_1,u_2)\,.
\end{array}
\label{TZA}
\eeq
This relation is known as the definition the {\em tetrahedral Zamolodchikov
  algebra}. It was introduced in \cite{Korepanov:1993}. 
The coefficients ${\cal
  S}_{d\,e\,f}^{a\,b\,c}(u_1,u_2,u_3)$ can be considered as the structure
constants of the algebra. They satisfy the tetrahedron equations
\eqref{TE3}, which 
play the role of the associativity condition for the defining
relations \eqref{TZA}. This relationship is analogues to that for 
the Zamolodchikov-Faddeev algebra \cite{Kulish:1981} 
where the Yang-Baxter equation
plays the role of the associativity condition.

In \cite{Korepanov:1993} Korepanov constructed a representation of the
the algebra \eqref{TZA}. He essentially postulated the expressions
\eqref{Ra-def}, without any connection to the 3D Zamolodchikov
model and then calculated the coefficients ${\cal
  S}_{d\,e\,f}^{a\,b\,c}(u_1,u_2,u_3)$, solving the linear equations
\eqref{TZA}. He used the normalization, where $\xi_1=\xi_2=\xi_3=1$,
so his expression for $\Sc_{123}$ (given before Eq.(2.32) in
\cite{Korepanov:1993}) contains square roots and, upon a change
of variables, exactly coincides with our expression for 
$\Su(\t_1,\t_2,\t_3)$, given by \eqref{weights2}. More precisely,
Korepanov used the variables $\l_1,\l_2,\l_3$ (which exacly correspond
to our variables $u_1,u_2,u_3$, respectively) and the 
variables $\varphi_1,\varphi_2,\varphi_3$, defined by Eq.(2.32) of
\cite{Korepanov:1993},
\beq
\tanh\varphi_i=\frac{1-\cd(2u_i)}{1+\cd(2u_i)},\quad i=1,2,3.
\eeq
Using addition theorems for elliptic and trigonometric functions it is
not difficult to show that 
\beq
g(u_j,u_k)=\tanh(\varphi_j-\varphi_k),\qquad j,k=1,2,3.
\eeq
where $g(x,y)$ is given by \eqref{g-def}. Then it follows from
\eqref{tan123} that
%% \beq
%% \tan\frac{\theta_1}{2}=-i\tanh(\varphi_1-\varphi_2),
%% \quad \tan\frac{\theta_2}{2}=\frac{i}{\tanh(\varphi_1-\varphi_3)},
%% \quad \tan\frac{\theta_3}{2}=-i\tanh(\varphi_2-\varphi_3).
%% \eeq
%% and 
\beq
\t_1=2 i (\varphi_2-\varphi_1),\qquad
\t_2=\pi + 2 i (\varphi_1-\varphi_3),\qquad
\t_3=2 i (\varphi_3-\varphi_2)\,.
\eeq

Subsequently, Shiroishi \& Wadati \cite{ShWad95b} reproduced the
calculations of \cite{Korepanov:1993}. They worked in essentially
the same normalization and notations as this work and their expressions for the
matrix elements of $\Sc_{123}$ (given by their Eq.(5.4)) 
exactly coincide with our Eq.\eqref{Snew} with interchanged $u_1$ and
$u_2$. In addition, Eq.\eqref{TZA} was re-checked once again in
\cite{Hubbook} in the
trigonometric limit ($k=0$), see \S  12.A.1 in \cite{Hubbook}.

%\newpage

\section{Conclusion}
In this paper we have considered a special limit of the tetrahedron
equation in the 3D Zamolodchikov model, where one of its vertices goes
to infinity and the tetrahedron turns into an infinite prism. We
explicitly showed that all vertex weights in the tetrahedron equation
\eqref{TE2} 
in this case are meromorphic functions depending on three points on an elliptic
curve. The parameterisation of the weights are given by \eqref{uniform}
and \eqref{Snew}. As a byproduct we obtained a parameterisation of angles
of a triangular prism in terms of elliptic functions (a problem, which is,
certainly, too hard for school geometry, but could have taken a
comfortable place among exercises in the classic textbook by Whittaker
\& Watson \cite{WW}). The corresponding formulae are given in
\eqref{teta456} and \eqref{teta123}.
  
Next, we have shown that a further reduction of this, already special, case of 
the tetrahedron equation leads precisely to the ``tetrahedral
Zamolodchikov algebra'',
originally invented in \cite{Korepanov:1993} 
without any connection to the 3D Zamolodchikov model.  
Interestingly, this algebra (more precisely, its 
trigonometric variant, corresponding to $k=0$) has been used by 
Shiroishi \& Wadati 
\cite{ShWad95a} for a two-layer construction of the Shastry's solution
\cite{Shastry:1986}
of the Yang-Baxter equation connected to the Hubbard model \cite{Lieb:1968}.
Recently, the interest to this solution was renewed due its remarkable
appearance \cite{Beisert:2006qh, Arutyunov:2006yd,Martins:2007hb,Mitev:2012vt} 
in the problem of the 
AdS/CFT correspondence for ${\cal N}=4$ SUSY Yang-Mills theory
in four dimensions.

Actually, one of motivations for this work was to unravel an
algebraic origin of the Shastry's $R$-matrix, which so far has not
been included into the standard quantum group scheme \cite{Dri87}.
This $R$-matrix has long been suspected to have a hidden 3D structure.
Indeed, it has an evident two-layer structure implied by the
construction of \cite{ShWad95a}, mentioned above.  
Moreover, it does not possess
the ``difference property'', which is another indication of a 3D
origin. To enhance the last argument it is worth mentioning that the
chiral Potts model \cite{AuY87,BPY87}, which is the most notable example
among solved models without the difference property, is just a
two-layer case of the multi-state generalisation of the 3D
Zamolodchikov model
\cite{Bazhanov:1992jqa,Bazhanov:1990qk1,Date:1990bs}.

Our results suggest that the Shastry's $R$-matrix is closely related to the 
3D Zamolodchikov model, which was originally formulated as 
an exact 
relativistic factorized scattering theory of ``straight strings''
in 2+1 dimensional space-time. This 3D model has an exremely rich
algebrac structure, previously studied in
\cite{Baxter:1983qc,Baxter:1986phd,
Bazhanov:1992jqa,Bazhanov:1993j,KMS:1993,Sergeev:1999jpa,Sergeev:1995rt,
BaxterForrester:1985,BaxterQuispel:1990,
  BaxterBazhanov:1997,BoosMangazeev:1999}.  
It would be interesting to see how this connection will help to
understand real origins of the Shastry's $R$-matrix.
We postpone these considerations to a separate publication \cite{BMS13a}.

\section* {Acknowledgements}

We thank N.~Beisert and W.~Galleas for attracting our attention to 
the works \cite{ShWad95a,ShWad95b,Hubbook}. This work was partially
supported by the Australian Research Council.

\bibliography{total32m}

\bibliographystyle{utphys}

\end{document}